\documentclass[aps,prx,twocolumn,groupedaddress,showkeys,showpacs,superscriptaddress,floatfix]{revtex4-2}

\usepackage{amsmath}
\usepackage{amssymb}
\usepackage{graphicx}
\usepackage{color}
\usepackage{orcidlink}

\begin{document}


\title{Comment on ``Non-reciprocal topological solitons in active metamaterials"}



\author{Duilio De Santis\,\orcidlink{0000-0001-6501-9763}}
\email[]{duilio.desantis@unipa.it}
\affiliation{Dipartimento di Fisica e Chimica ``E.~Segr\`{e}", Group of Interdisciplinary Theoretical Physics, Università degli Studi di Palermo, I-90128 Palermo, Italy}

\author{Bernardo Spagnolo\,\orcidlink{0000-0002-6625-3989}}
\email[]{bernardo.spagnolo@unipa.it}
\affiliation{Dipartimento di Fisica e Chimica ``E.~Segr\`{e}", Group of Interdisciplinary Theoretical Physics, Università degli Studi di Palermo, I-90128 Palermo, Italy}
\affiliation{Radiophysics Department, Lobachevsky State University, 603950 Nizhniy Novgorod, Russia}

\author{Angelo Carollo\,\orcidlink{0000-0002-4402-2207}}
\email[]{angelo.carollo@unipa.it}
\affiliation{Dipartimento di Fisica e Chimica ``E.~Segr\`{e}", Group of Interdisciplinary Theoretical Physics, Università degli Studi di Palermo, I-90128 Palermo, Italy}

\author{Davide Valenti\,\orcidlink{0000-0001-5496-1518}}
\email[]{davide.valenti@unipa.it}
\affiliation{Dipartimento di Fisica e Chimica ``E.~Segr\`{e}", Group of Interdisciplinary Theoretical Physics, Università degli Studi di Palermo, I-90128 Palermo, Italy}

\author{Claudio Guarcello\,\orcidlink{0000-0002-3683-2509}} 
\email[]{cguarcello@unisa.it}
\affiliation{Dipartimento di Fisica ``E.~R.~Caianiello", Università degli Studi di Salerno, I-84084 Fisciano, Salerno, Italy}
\affiliation{INFN, Sezione di Napoli, Gruppo Collegato di Salerno - Complesso Universitario di Monte S. Angelo, I-80126 Napoli, Italy}


\date{\today}



\maketitle



In Ref.~\cite{Veenstra_2024}, for an analytical understanding of the system under consideration, the authors derive an ordinary differential equation for the sine-Gordon (SG) (anti)soliton velocity, with the perturbation theory in the adiabatic approximation, via the inverse scattering transform formalism, see Eq.~(3) and section ``Theoretical methods'' in their work.

Here we note that Eq.~(3) in Ref.~\cite{Veenstra_2024} for the (anti)soliton velocity also follows from an energy balance approach~\cite{McLaughlin_1978_PRA, Scott_2003}. We consider the non-reciprocal SG equation for the field ${ \phi(x, t) }$
\begin{equation}
\label{eqn:1}
\frac{\partial^2 \phi}{\partial t^2} - \frac{\partial^2 \phi}{\partial x^2} + \sin \phi = - \eta \frac{\partial \phi}{\partial x} - \Gamma \frac{\partial \phi}{\partial t} ,
\end{equation}
with ${ \eta }$ and ${ \Gamma }$ being, respectively, the non-reciprocal driving strength and the dissipation coefficient. In the absence of perturbations, i.e., for ${ \eta = \Gamma = 0 }$, the energy
\begin{equation}
\label{eqn:2}
H = \int dx \left[ \frac{1}{2} \left( \frac{\partial \phi}{\partial t} \right)^2 + \frac{1}{2} \left( \frac{\partial \phi}{\partial x} \right)^2 + \left( 1 - \cos \phi \right) \right] 
\end{equation}
is conserved. In our case, we have that
\begin{equation}
\label{eqn:3}
\frac{dH}{dt} = \int dx \left[ \left( - \eta \frac{\partial \phi}{\partial x} - \Gamma \frac{\partial \phi}{\partial t} \right) \frac{\partial \phi}{\partial t} \right] ,
\end{equation}
due to the energy input (${ \eta }$ -- non-reciprocal driving) and the loss (${ \Gamma }$ -- damping). The assumption of the perturbation theory is that, if ${ \eta }$ and ${ \Gamma }$ coefficients are small, the main effect on a topological soliton is to alter its velocity, making the latter a slowly time-dependent quantity.

Owing to the relativistic structure of the unperturbed SG equation, the energy of a soliton travelling with velocity ${ v < 1 }$ is
\begin{equation}
\label{eqn:4}
H = \frac{8}{\sqrt{1 - v^2}} .
\end{equation}
We can compute the LHS of Eq.~\eqref{eqn:3} by means of the latter formula, allowing for a slow time-variation of the parameter ${ v }$. The integral of Eq.~\eqref{eqn:3}'s RHS is then evaluated via the single soliton solution of the unperturbed SG equation, keeping ${ v }$ fixed. It readily follows that
\begin{equation}
\label{eqn:5}
\frac{dv}{dt} = - \left(1 - v^2 \right) \left( \Gamma v - \eta \right) ,
\end{equation}
in accordance with the expression given in Ref.~\cite{Veenstra_2024}.

By also including a constant force term ${ f }$ in Eq.~\eqref{eqn:1}, the previous steps result in
\begin{equation}
\label{eqn:6}
\frac{dv}{dt} = \frac{\pi f}{4} \left( 1 - v^2 \right)^{3/2} - \left(1 - v^2 \right) \left( \Gamma v - \eta \right) ,
\end{equation}
which is also used in Ref.~\cite{Veenstra_2024}.

\begin{acknowledgments}

We are grateful to the authors of Ref.~\cite{Veenstra_2024} for the feedback they provided on the present topic.

\end{acknowledgments}



%

\end{document}